\documentclass[floats,floatfix,showpacs,amssymb,prd,twocolumn,superscriptaddress,nofootinbib,nolongbibliography,reprint]{revtex4-2}

\usepackage{amssymb, amsmath, verbatim, mathtools, needspace, enumitem, etoolbox, graphicx, physics, microtype, afterpage, bigints, gensymb, tabularx, xspace}

\usepackage[dvipsnames, usenames]{xcolor}
\definecolor{linkcolor}{rgb}{0.0,0.3,0.5}
\definecolor{dodgerblue}{HTML}{1E90FF}
\usepackage[unicode, colorlinks=true, linkcolor=linkcolor, citecolor=linkcolor, filecolor=linkcolor,urlcolor=linkcolor, pdfusetitle]{hyperref}
\usepackage[all]{hypcap}
\usepackage[T1]{fontenc}
\usepackage[utf8]{inputenc}
\usepackage{orcidlink}
\usepackage{natbib}
\usepackage{url}
\usepackage{cleveref,multirow}
\usepackage[caption=false]{subfig} 
\usepackage[normalem]{ulem} %

\newcommand{\ssim}{\mathchar"5218\relax\,}

\makeatletter
\newcommand*{\balancecolsandclearpage}{\close@column@grid \cleardoublepage \twocolumngrid}
\makeatother

\usepackage{lipsum}

\def\apj{{Astrophys.~J.}}
\def\apjl{{Astrophys.~J.~Lett.}}

\def\prd{{Phys.~Rev.~D}}

\def\prl{{Phys.~Rev.~Lett.}}
\def\prr{{Phys.~Rev.~Res.}}

\def\lrr{{Living Rev. Relativ.}}
\def\cqg{{Classical~Quant.~Grav.}}

\def\surname{{NRSur7dq4Remnant}\xspace}
\def\newname{{NRSur7dq4EmriRemnant}\xspace}

\newcommand{\bham}{\affiliation{School of Physics and Astronomy \& Institute for Gravitational Wave Astronomy, University of Birmingham,\\ Birmingham, B15 2TT, United Kingdom}}

\newcommand{\milan}{\affiliation{Dipartimento di Fisica ``G. Occhialini'', Universit\'a degli Studi di Milano-Bicocca, Piazza della Scienza 3, 20126 Milano, Italy}}
\newcommand{\infn}{\affiliation{INFN, Sezione di Milano-Bicocca, Piazza della Scienza 3, 20126 Milano, Italy}}
\newcommand{\aei}{\affiliation{Max Planck Institute for Gravitational Physics (Albert Einstein Institute), Am M\"uhlenberg 1, Potsdam 14476, Germany}}
\newcommand{\dartmouth}{\affiliation{Department of Mathematics, Center for Scientific Computing and Data Science Research,\\ University of Massachusetts, Dartmouth, MA 02747, USA}}
\newcommand{\csu}{\affiliation{Nicholas and Lee Begovich Center for Gravitational-Wave Physics and Astronomy, California State University, Fullerton, Fullerton, California 92831, USA}}
\newcommand{\cornell}{\affiliation{Cornell Center for Astrophysics and Planetary Science, Cornell University, Ithaca, New York 14853, USA}}
\newcommand{\cornellPhys}{\affiliation{Department of Physics, Cornell University, Ithaca, New York 14853, USA}}
\newcommand{\caltech}{\affiliation{TAPIR 350-17, California Institute of Technology, 1200 E California Boulevard, Pasadena, California 91125, USA}}
\newcommand{\tata}{\affiliation{International Centre for Theoretical Sciences, Tata Institute of Fundamental Research, Bangalore 560089, India}}

\DeclareUnicodeCharacter{2212}{-}
\DeclareMathOperator{\sgn}{sgn}
\DeclareMathOperator{\arctanh}{arctanh}

\begin{document}

\title{Extending black-hole remnant surrogate models
to extreme mass ratios}

\author{Matteo Boschini$\,$\orcidlink{0009-0002-5682-1871}}
\email{m.boschini1@campus.unimib.it}

\milan
\aei

\author{Davide Gerosa$\,$\orcidlink{0000-0002-0933-3579}}

\milan \infn \bham

\author{Vijay Varma$\,$\orcidlink{0000-0002-9994-1761}}

\aei \dartmouth

\author{\\\medskip Cristóbal Armaza$\,$\orcidlink{0000-0002-1791-0743}}
\cornell

\author{Michael Boyle$\,$\orcidlink{0000-0002-5075-5116}}
\cornell

\author{Marceline S. Bonilla$\,$\orcidlink{0000-0003-4502-528X}}
\csu

\author{Andrea Ceja$\,$\orcidlink{0000-0002-1681-7299}}
\csu

\author{Yitian Chen$\,$\orcidlink{0000-0002-8664-9702}}
\cornell

\author{\\Nils Deppe$\,$\orcidlink{0000-0003-4557-4115}}
\caltech \cornellPhys

\author{Matthew Giesler$\,$\orcidlink{0000-0003-2300-893X}}
\cornell 

\author{Lawrence E. Kidder$\,$\orcidlink{0000-0001-5392-7342}}
\cornell

\author{Prayush Kumar$\,$\orcidlink{0000-0001-5523-4603}}
\tata

\author{Guillermo Lara$\,$\orcidlink{0000-0001-9461-6292}}
\aei

\author{\\Oliver Long$\,$\orcidlink{0000-0002-3897-9272}}
\aei

\author{Sizheng Ma$\,$\orcidlink{0000-0002-4645-453X}}
\caltech

\author{Keefe Mitman$\,$\orcidlink{0000-0003-0276-3856}}
\caltech

\author{Peter James Nee$\,$\orcidlink{0000-0002-2362-5420}}
\aei

\author{Harald P. Pfeiffer$\,$\orcidlink{0000-0001-9288-519X}}
\aei

\author{\\Antoni Ramos-Buades$\,$\orcidlink{0000-0002-6874-7421}}
\aei

\author{Mark A. Scheel$\,$\orcidlink{0000-0001-6656-9134}}
\caltech

\author{Nils L. Vu$\,$\orcidlink{0000-0002-5767-3949}}
\caltech

\author{Jooheon Yoo$\,$\orcidlink{0000-0002-3251-0924}}
\cornell

\pacs{}

\date{\today}

\begin{abstract}

Numerical-relativity surrogate models for both black-hole merger waveforms and
remnants have emerged as important tools in gravitational-wave astronomy. While
producing very accurate predictions, their  applicability is limited to the
region of the parameter space where numerical-relativity simulations are
available and  computationally feasible. Notably, this excludes extreme mass
ratios. We present a %
machine-learning approach to extend the validity of existing and
future numerical-relativity surrogate models toward the test-particle limit,
targeting in particular the mass and spin of post-merger black-hole remnants.
Our model is trained on both numerical-relativity simulations at comparable
masses and analytical predictions at extreme mass ratios. We extend the
gaussian-process-regression model \surname, %
validate its performance via cross validation, and test
its accuracy against %
additional numerical-relativity runs.
Our fit, which we dub \newname, reaches an accuracy that is comparable to or higher than that of existing
remnant models while providing robust predictions for arbitrary mass ratios.
\end{abstract}

\maketitle

\section{Introduction}

Accurate modeling of merging black-hole (BH) binaries is crucial to both
understanding the two-body problem in General Relativity (GR) and
characterizing gravitational-wave (GW) observations.  Numerical relativity (NR)
simulations~\cite{2015CQGra..32l4011S} are, at present, the most accurate
approach to capture the dynamics of merging BHs as well as the emitted
gravitational signal. Despite the great computational cost of NR, several
groups are compiling extensive catalogs of simulations with ever-increasing
parameter-space coverage~\cite{2016CQGra..33t4001J,
2019CQGra..36s5006B, 2022PhRvD.105l4010H}.

NR simulations are a crucial ingredient to calibrate approximate and
fast-to-evaluate emulators for data analysis and astrophysical modeling (for reviews see Refs.~\cite{2012CQGra..29l4002O, 2014LRR....17....2B,2016LNP...905..273D,2022hgwa.bookE..38P,2022LRR....25....2T}). %
Among these, surrogate models are assuming a prominent role. Compared to other
approaches, surrogates interpolate the underlying NR simulations in a purely
data-driven fashion without ansatzes based on post-Newtonian
(PN) and BH-perturbation theory. %
NR surrogates for both waveforms 
~\cite{2017PhRvD..96b4058B, 2019PhRvR...1c3015V, 2021PhRvD.103f4022I, 2022PhRvD.106d4001Y, 2023arXiv230603148Y} 
and post-merger remnants
~\cite{2018PhRvD..97j4049G, 2019PhRvR...1c3015V,2019PhRvL.122a1101V,
2020CQGra..37v5005R,2021PhRvD.103f4022I,2023arXiv230107215I, 2020CQGra..37m5005H} are now widely available and routinely used in GW astronomy.

While generically more accurate than other approximants, the main limitation of
surrogate models is their applicability regime, which is largely restricted to
the region of the parameter space used for training. Notably, the cost of NR
simulations increases for BH binaries with unequal masses, ultimately becoming
prohibitive as one approaches the extreme mass-ratio inspiral (EMRI) limit. At
present, NR surrogate models for both waveforms and remnants extrapolate rather
poorly in this regime.

{\surname}~\cite{2019PhRvR...1c3015V} is the state-of-the-art surrogate model
for the mass, spin, and 
recoil
velocity (or kick) of the post-merger remnant.
The training set is composed of 1528 NR simulations
from the Simulating eXtreme Spacetimes (SXS) %
catalog~\cite{2019CQGra..36s5006B} which are interpolated with
Gaussian Process Regression (GPR)~\cite{RasmussenW06}.
The model covers the 7-dimensional parameter space spanned by %
quasicircular binaries:
the mass
ratio $q=m_1/m_2$ (here defined to be $\geq1$ for consistency with previous work),
and the dimensionless spins of the two BHs $\boldsymbol{\chi}_{1,2}$ (where
labels 1 and 2 refer to the heavier and lighter BH, respectively), thus
including spin precession. The model predicts remnant mass $m_{\rm f}/M$ (where $M=m_1+m_2$ is the total mass), spin vector $\boldsymbol{\chi}_{\rm f}$,
and recoil vector $\boldsymbol{v}_{\rm f}$.
The training dataset is restricted to binaries with $q \leq 4$ and
${\chi}_{1,2} \leq 0.8$. While numerical extrapolation to the high-spin limit and up to mass ratios $q=6$
return reasonable results~\cite{2019PhRvR...1c3015V}, a naive extension of the
model toward the EMRI limit fails dramatically. For $q\gg 4$, \surname often
returns predictions that are wildly unphysical, such as remnant masses that are
greater than the total mass of the binary and/or spins that exceed the Kerr
limit.

In this paper, we develop a procedure to consistently extend NR surrogate
models for the remnant mass and remnant spin vector %
to the extreme mass-ratio region.
We make use of analytical limits for the energy and angular momentum carried
away by GWs to generate suitable training data in addition to the NR
simulations used in \surname. We identify the optimal set of tuning parameters
using cross-validation and compare our predictions against an additional test
set of NR simulations. Our fits, which we dub \newname, are as accurate as the existing ones in the
nominal NR training region while %
extending their validity to extreme mass
ratios with similar residuals.

This paper is organized as follows. In Sec.~\ref{sec:two} we write down the
EMRI limit for $m_{\rm f}$ and ${\boldsymbol\chi}_{\rm f}$. In
Sec.~\ref{sec:three} we presents training and validation of our augmented
model. In Sec.~\ref{sec:four} we test our results against additional NR
simulations. In Sec.~\ref{sec:five} we investigate the computational costs of
our extended models. In Sec.~\ref{sec:six} draw our conclusions. We use
geometric unit where $c=G=1$.

\section{The extreme-mass-ratio limit}
\label{sec:two}

Our procedure relies on analytical expressions for the mass and spin of the
post-merger BH as a function of the binary parameters  that are valid in the
EMRI limit.

\subsection{Remnant mass}
\label{remnantmass}

The remnant mass is equal to the difference between the binary total mass and
the energy radiated in GWs. 
We assume that the secondary BH inspirals adiabatically till the innermost
stable circular orbit (ISCO)
and neglect energy dissipation from the subsequent plunge onward. This is
appropriate because, even though the energy loss rate might be large, the
transition from ISCO to merger happens in a short
time~\cite{1996PhRvD..53.4319K}. 

In this approximation, the final mass is given by  (e.g. Ref.~\cite{2012ApJ...758...63B})%
\begin{align}
	\label{eq:mf_foq}
 \frac{m_{\rm f}}{M} = 1-\frac{1}{q}\big[1-E_{\rm ISCO}(\chi_1\cos\theta_1)\big] + \mathcal O\left(\frac{1}{q^2}\right) \,,
\end{align}
where $\theta_1= \arccos \hat{\boldsymbol{\chi}}_1\cdot
\hat{\boldsymbol{L}}_{\rm ISCO}$ is the angle between the spin of the primary
and the orbital angular momentum at the ISCO.
The separation and energy at the ISCO %
in units of $m_1$ and $m_2$, respectively, 
are~\cite{1972ApJ...178..347B} %
\begin{align}
	& E_{\rm ISCO}(\chi) = \sqrt{1-\frac{2}{3 \, r_{\rm ISCO}(\chi)}} \,, \\
	\label{eq:r_ISCO}
	& r_{\rm ISCO}(\chi) = 3+Z_2-\sgn(\chi)\sqrt{(3-Z_1)(3+Z_1+2Z_2)} \,,\\
	& Z_1 = 1+(1-\chi^2)^{1/3}\bigl[(1-\chi)^{1/3}+(1+\chi)^{1/3}\bigr] \,,  \\
	& Z_2 = \sqrt{3\chi^2+2Z_1^2} \,. 
\end{align}
The additional terms of  $\mathcal{O}(1/q^2)$ in Eq.~(\ref{eq:mf_foq}) account
for the modification of the ISCO due to the non-zero mass of the secondary
BH~\cite{2009PhRvL.102s1101B, 2012PhRvL.108m1103L} as well as deviations from
the adiabatic inspiral near plunge~\cite{2011PhRvD..83j4011K}.

The expression above is equivalent\footnote{Note that
Refs.~\cite{2012ApJ...758...63B,2016ApJ...825L..19H} adopt a convention where
$q\leq 1$ while here we use $q\geq 1$ for compatibility with
Ref.~\cite{2019PhRvR...1c3015V}.} to that of
\citeauthor{2012ApJ...758...63B}~\cite{2012ApJ...758...63B} to first order in
$1/q$. Those authors then augment the EMRI ansatz with coefficients that are
calibrated on the NR simulations available at the time. They also include
information on the spin of the secondary by inserting a suitable weighted
combinations of the two spins into the expression for the ISCO energy. In the
interest of avoiding uncontrolled systematics from a different NR dataset, here
we implement the simple expression from Eq.~(\ref{eq:mf_foq}).

\subsection{Remnant spin} 
\label{remnantspin}

We model the spin of the remnant as the sum of the two BH spins and orbital
angular momentum at the ISCO: 
\begin{equation}
	\label{eq:start_Chif}
	m_{\rm f}^2 \boldsymbol{\chi}_{\rm f} = m_1^2  \boldsymbol{\chi}_1 +m_2^2 \boldsymbol{\chi}_2  + m_1 m_2\boldsymbol{L}_{\rm ISCO}\,,
\end{equation}
thus neglecting once more the contribution to the GW flux from the plunge
onward. Taylor expanding Eq.~(\ref{eq:start_Chif}) to first order in $1/q$
yields
\begin{align}
	\label{eq:chif_foq}
	\boldsymbol{\chi}_{\rm f} = \boldsymbol{\chi}_1 &+ \frac{1}{q}\big[\boldsymbol{L}_{\rm ISCO}(\chi_1\cos\theta_1) \notag\\
	&
	 - 2\boldsymbol{\chi}_1E_{\rm ISCO}(\chi_1\cos\theta_1)\big] 
	+ \mathcal O\left(\frac{1}{q^2}\right) \,,
\end{align}
where~\cite{1972ApJ...178..347B}
\begin{equation}
	L_{\rm ISCO}(\chi) = \frac{2}{3\sqrt{3}}\left[1+2\sqrt{3 r_{\rm ISCO}(\chi)-2}\right] \,.
\end{equation}
is the orbital angular momentum at the ISCO in units of $m_1m_2$.

This expression agrees with that presented by
\citeauthor{2016ApJ...825L..19H}~\cite{2016ApJ...825L..19H} (see also
Ref.~\cite{2009ApJ...704L..40B}) as long as terms of $\mathcal{O}(1/q^2)$ are
neglected. Much like the mass case described above, we opt for a cleaner ansatz
that only contains information on the test-particle limit without further
assumption on the role of the secondary spin and/or calibrated coefficients.

\section{Model training}
\label{sec:three}

Our machine-learning model is trained on a joint dataset
made of NR simulations covering the comparable-mass regime and analytical
predictions in the EMRI limit.  We wish to provide a  representation of the
mapping between the 7-dimensional space  $(q, \boldsymbol{\chi}_1,
\boldsymbol{\chi}_2)$ and  the 4-dimensional space $(m_{\rm
f},\boldsymbol{\chi}_{\rm f})$.

\subsection{Constructing the training dataset}

The NR training set is made of the same 1528 SXS simulations used in
Ref.~\cite{2019PhRvR...1c3015V}; see therein for details and further
references. %

For the EMRI datapoints, we consider $n_{\rm EMRI}$ binaries with mass ratios
distributed between $q_{\rm min}$ and $q_{\rm max}$. %
We distribute spin
magnitudes $\chi_{1,2}$ uniformly in $[0,1]$, polar angles (measured from the
orbital angular momentum) $\theta_{1,2}$ uniformly in $[0, \pi]$, and azimuthal
angles (measured in the orbital plane) $\phi_{1,2}$ uniformly in $[0, 2\pi]$.
We sample $\theta_{1,2}$ uniformly (and not uniformly in cosine) in
order to increase accuracy for aligned spins; this is a configuration that is astrophysically relevant and is also overrepresented in the test set, cf. Sec~\ref{sec:four}. %
We sample the mass ratio from a log-uniform distribution, which
is the same %
input passed to the GPR algorithm~\cite{2019PhRvR...1c3015V}.
Note that the EMRI training set includes values of the spin of the secondary,
even though this is ignored at $\mathcal{O}(1/q)$,  cf. Secs.~\ref{sec:two}

All input quantities in \surname are specified at
a time $t = −100M$ before merger~\cite{2019PhRvR...1c3015V}
in a frame where the $z$-axis is along the orbital angular
momentum, the $x$-axis is along the line of separation
from the less to the more massive BH, and the $y$-axis completes the
right-handed triad. We refer to this frame as the ``{wave frame}'' at $t=-100 M$ and note it is defined using the waveform at future
null infinity~\cite{2019PhRvR...1c3015V}.
The choice of $t = -100M$ is
well suited for modeling the merger remnant for comparable-mass binaries, as this time falls within $2-4$
GW cycles before the peak GW amplitude for comparable-mass binaries~\cite{2022PhRvD.105b4045V}. We use this
same reference frame for the NR binaries of the training set %
for the new
model \newname.

For the additional EMRI inputs, %
we instead use the wave frame
at the ISCO. Defining $t = -100M$ before peak-strain amplitude is
challenging for EMRIs as it would require a detailed spin-evolution procedure which is valid in this regime. On
the other hand, the ISCO %
naturally indicates a point near
merger where conservation laws can be applied (Sec.~\ref{sec:two}), which implies the resulting datapoints can be (at least approximately) put together with the NR dataset at $t = -100M$.
In practice this means that when evaluating the EMRI limits of $m_{\rm f}$ and
$\boldsymbol{\chi}_{\rm f}$ we are
neglecting the evolution of the angular momenta between $t = -100M$ and the
ISCO for our EMRI data, which is a reasonable assumption~\cite{1994PhRvD..49.6274A}.

The quantities returned by our model would then also be defined in the wave
frame at $t=-100 M$ in the NR regime and in the wave frame at ISCO in the EMRI
regime, with a transition between the two frames in the intermediate regimes.
This is reasonable for this work as the remnant mass and spin are relatively insensitive to small changes in the spins near merger. A more careful approach
may be necessary to model the remnant kick, which instead highly depends on the
spin directions before merger \cite{2018PhRvD..97j4049G}.

\subsection{Fit optimization}
\label{outputopt}

Training is performed using GPR~\cite{RasmussenW06}. We use the same kernel choices and parameter settings of Refs.~\cite{2019PhRvL.122a1101V,2019PhRvR...1c3015V}, which appear to work well also in this case.

When estimating the final mass, we enforce the physical constraint $m_{\rm f}/ M \leq 1$ by appropriately transforming the input data. 
In particular, we fit for 
$\arctanh(m_{\rm f}/M)$, thus mapping  the interval $[0,1)$ to %
$[0, \infty)$.
When evaluating the trained model, we then reverse the operation %
($\tanh$: $[0, \infty) \to[0,1)$), which ensures that the remnant mass is properly limited. Furthermore, we empirically find that this choice enhances the overall performance of the model because it simplifies the fitting process: the function tanh resembles the expected behavior of $m_{\rm f}/M$ as a function of $q$, so the training data become closer to linear in that dimension. %

As for the remnant spin, we use the GPR strategy of
Ref.~\cite{2019PhRvR...1c3015V} which requires the cartesian components of
$\boldsymbol{\chi}_{\rm f}$ and fits them separately. In the following, we report results in terms of 
magnitude %
$\chi_{\rm f}$, polar angle $\theta_{\rm f}$
measured from the pre-merger binary orbital angular momentum and azimuthal
angle $\phi_{\rm f}$ measured in the pre-merger orbital plane. For simplicity, we  focus on the first two of these quantities because they are more
relevant for astrophysical applications. 

For extremely large mass ratios $q \gg q_{\rm max}$, the
 GPR error on $\chi_{\rm f}$ becomes comparable to the residual, which  implies we are better off
assuming the EMRI limit at the evaluation stage instead of asking GPR to learn
it from the training simulations. We use a squared sinusoidal transition
function to smoothly connect fit outcomes and the exact EMRI limit. For mass ratios $q_{\rm max} < q < 2 q_{\rm max}$ we correct each cartesian component $\chi_{\rm f, i }$ for $i=\{x,y,z\}$ using the affine transformation
\begin{equation}
\frac{\chi_{\rm f, i }(q) -  \chi_{\rm f,i}^{\rm fit}(q_{\rm max})}{{\chi_{\rm f,i}^{\rm EMRI}(2q_{\rm max}) - \chi_{\rm f,i}^{\rm fit}(q_{\rm max})}} = \sin^2 \left(\pi \frac{q-q_{\rm max}}{2 q_{\rm max}}\right)\,,
\end{equation}
where superscripts ``fit'' and ``EMRI'' indicate GPR and analytical predictions, respectively. Note we only correct along the $q$ direction of the fit. %
This output corrections is not
necessary for the remnant mass because the zero-th order EMRI limit
$m_{\rm f}/M\to 1$ is already imposed by the tanh transformation described
above.

\subsection{Cross validation and parameter tuning}
\label{crossval}

\begin{figure*}[]
	\centering
	\subfloat{\includegraphics[height=.3\textwidth]{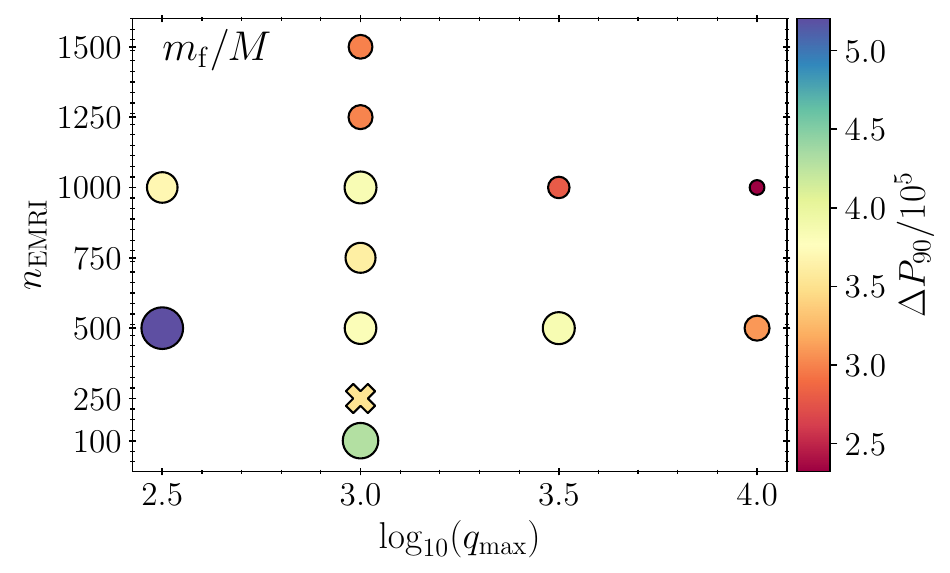}}
	\qquad
	\subfloat{\includegraphics[height=.3\textwidth]{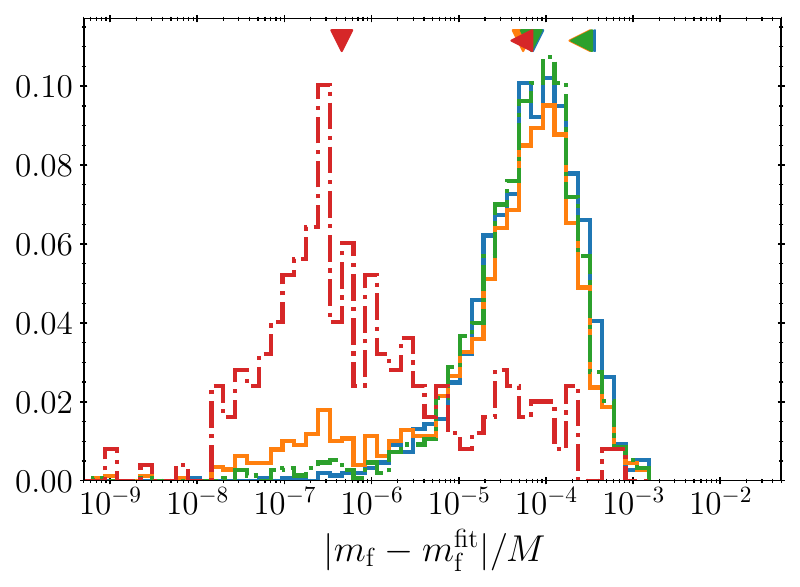}} \\
	\subfloat{\includegraphics[height=.3\textwidth]{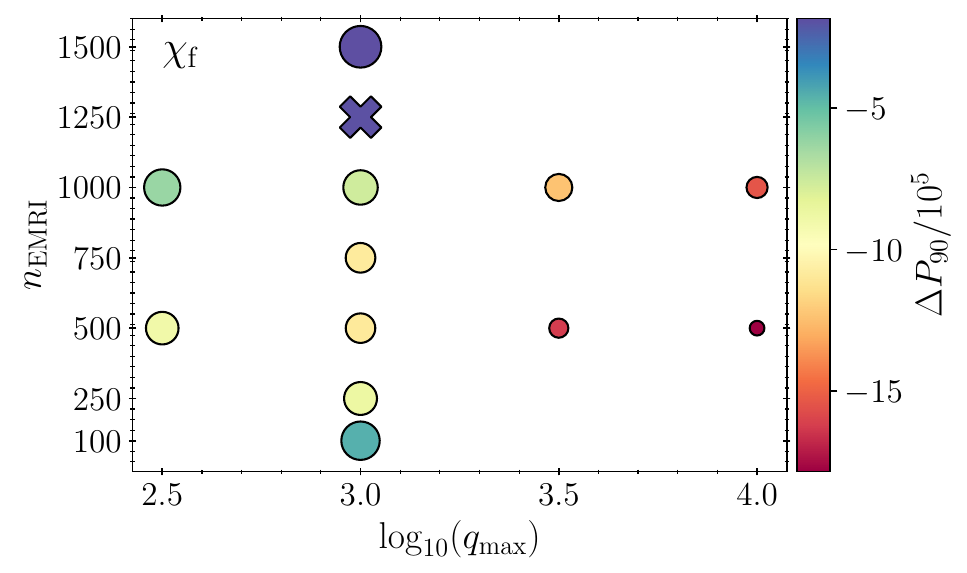}}
	\qquad
	\subfloat{\includegraphics[height=.3\textwidth]{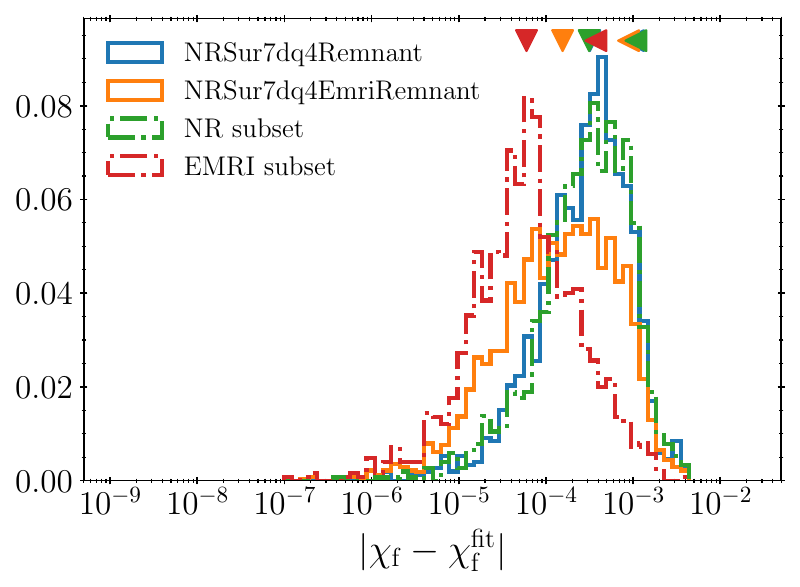}} \\
	\subfloat{\includegraphics[height=.3\textwidth]{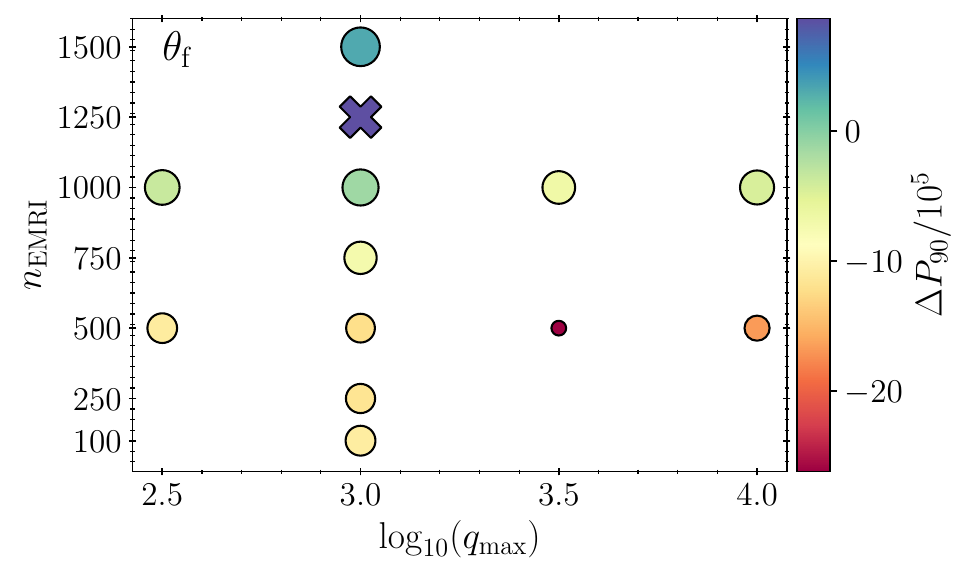}}
	\qquad
	\subfloat{\includegraphics[height=.3\textwidth]{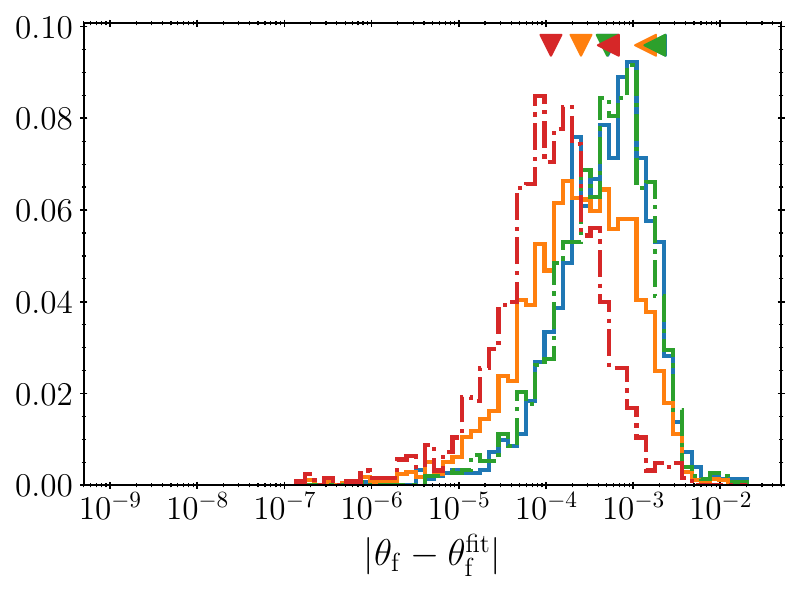}}
\caption{
Outcome of a 20-fold cross validation on the combined NR and EMRI training
dataset. Top, middle, and bottom panels show results for the final mass $m_{\rm
f}/M$, the final spin magnitude $\chi_{\rm f}$, and the polar angle
$\theta_{\rm f}$. The left panels show the 90$^{\rm th}$ percentile
$\Delta{P}_{90}$ of the validation residuals varying over the number of added
EMRI systems $n_{\rm EMRI}$ and the upper bound of the EMRI training region
$q_{\rm max}$ while keeping the lower bound fixed to $q_{\rm min} = 10^2$. The
mass (spin) fit with $n_{\rm EMRI} = 250$ ($n_{\rm EMRI}= 1250$) and $q_{\rm
max} = 1000$ are marked with crosses  and selected as optimal, see
Sec.~\ref{sec:three}. The right panels show the full distribution of the
validation residuals for these cases. The validation results computed on the
full training dataset are shown in orange; these are then broken down by
selecting only the NR (green) and EMRI (red) samples. Residuals from the
\surname model as published in Ref.~\cite{2019PhRvR...1c3015V} are shown in
blue. Left- (down-) pointing triangles indicate the 90$^{\rm th}$ percentile
(median) of the corresponding distributions.
}
\label{fig:hyp_span}
\end{figure*}

When adding EMRI training points, we set the lower end of the mass-ratio
interval to $q_{\rm min}=100$. We verified numerically that at this value of
the mass ratio the discrepancy between our first-order relations from
Sec.~\ref{sec:two} and the analytic formulas from
Refs.~\cite{2012ApJ...758...63B,2016ApJ...825L..19H} is of the same order of
the $95^{\rm th}$ percentile residuals in the \surname
model \cite{2019PhRvR...1c3015V}, see their Fig.~7.  We optimize the number of EMRI
binaries $n_{\rm EMRI}$  and the upper edge of the mass ratio regime $q_{\rm
max}$ with a grid search. For each of the 13 parameter-space locations shown in
the left panels of Fig.~\ref{fig:hyp_span}, we perform a %
 20-fold cross
validation~\cite{10.1111/j.2517-6161.1974.tb00994.x} and record the 90$^{\rm
th}$ percentile of the absolute differences $\Delta P_{90}$ between the true
and fitted values of $m_{\rm f}/M$, $\chi_{\rm f}$, and $\theta_{\rm f}$.  For
this hyperparameters search, the variations on the final mass are of
$\mathcal{O}(10^{-5})$ while those on the final spin are of $\mathcal{O}(10^{-4})$.

{An optimal choice of the hyperparameters requires a balance between accuracy and computational costs.}
Considering the grid search results, the test performance (Sec.~\ref{sec:four}), and the
computational cost (Sec.~\ref{sec:five}), we set $q_{\rm max} =10^3$ and
$n_{\rm EMRI}=250, 1250$ for the $m_{\rm f}$ and $\boldsymbol{\chi}_{\rm f}$
fit, respectively. 
{The corresponding models show high accuracy in both validation and test with a limited increase of computational evaluation time compared to \surname.}
Although some of the alternative fits from
Fig.~\ref{fig:hyp_span} returns marginally better results (see e.g. the $q_{\rm
max} =10^{2.5}$ case for $m_{\rm f}$), we believe setting the same mass-ratio
interval for both fits (mass and spin) ensures a smoother procedure in light of
future retraining with more NR simulations, when available. It is also worth
stressing that the difference between these two models $\Delta P_{90}\sim
\mathcal{O}(10^{-5})$ is an order of magnitude smaller than the test errors,
see Sec.~\ref{sec:four}.

The right panel of Fig.~\ref{fig:hyp_span} shows the distributions of the
validation residuals for these optimal fits. Considering the full validation
estimates, our extended fits \newname have performances that are comparable to those of
\surname from Ref.~\cite{2019PhRvR...1c3015V} (see their Fig. 7) while
extending its validity to the extreme-mass ratio regime. This is made evident
by separating the two subsets of EMRI and NR training data: the NR data are
fitted as accurately as \surname and validation errors for the EMRI datapoints
are smaller by several orders of magnitude.

\section{Model performance}
\label{sec:four}

The cross-validation procedure presented above is an internal consistency
check, where the model is tested in the same mass-ratio region used for
training. More ambitiously, we also test our models against an independent NR
dataset. Our test set is made of 1228 NR simulations from the SXS group. These were recently
performed for building a new waveform approximant~\cite{inprep} and will be released publicly together with that model. Our test dataset
contains numerous systems with $4<q<100$ 
that allows us to test the behavior of the extended fits in
between the NR and EMRI training regions. We compare the performance of our
extended model \newname against  those from \surname as published in
Ref.~\cite{2019PhRvR...1c3015V} as well as the analytic formulas from
Refs.~\cite{2012ApJ...758...63B,2016ApJ...825L..19H} (hereafter ``HBMR'') with $n_M=3$ and $n_J=4$. In the following, we do not show HBMR predictions for $\theta_{\rm f}$ because that quantity is not readily accessible from their fits as a function of $q, \boldsymbol{\chi}_1,$ and $\boldsymbol{\chi}_{2}$. NR surrogates have been extensively tested against other fits used in GW astronomy \cite{2017PhRvD..95f4024J, 2017PhRvD..95b4037H, DCC_remnant} with similar results, cf. Ref.~\cite{2019PhRvL.122a1101V}.
\subsection{Test example}

\begin{figure*}[]
	\centering
	\includegraphics[width=0.92\textwidth]{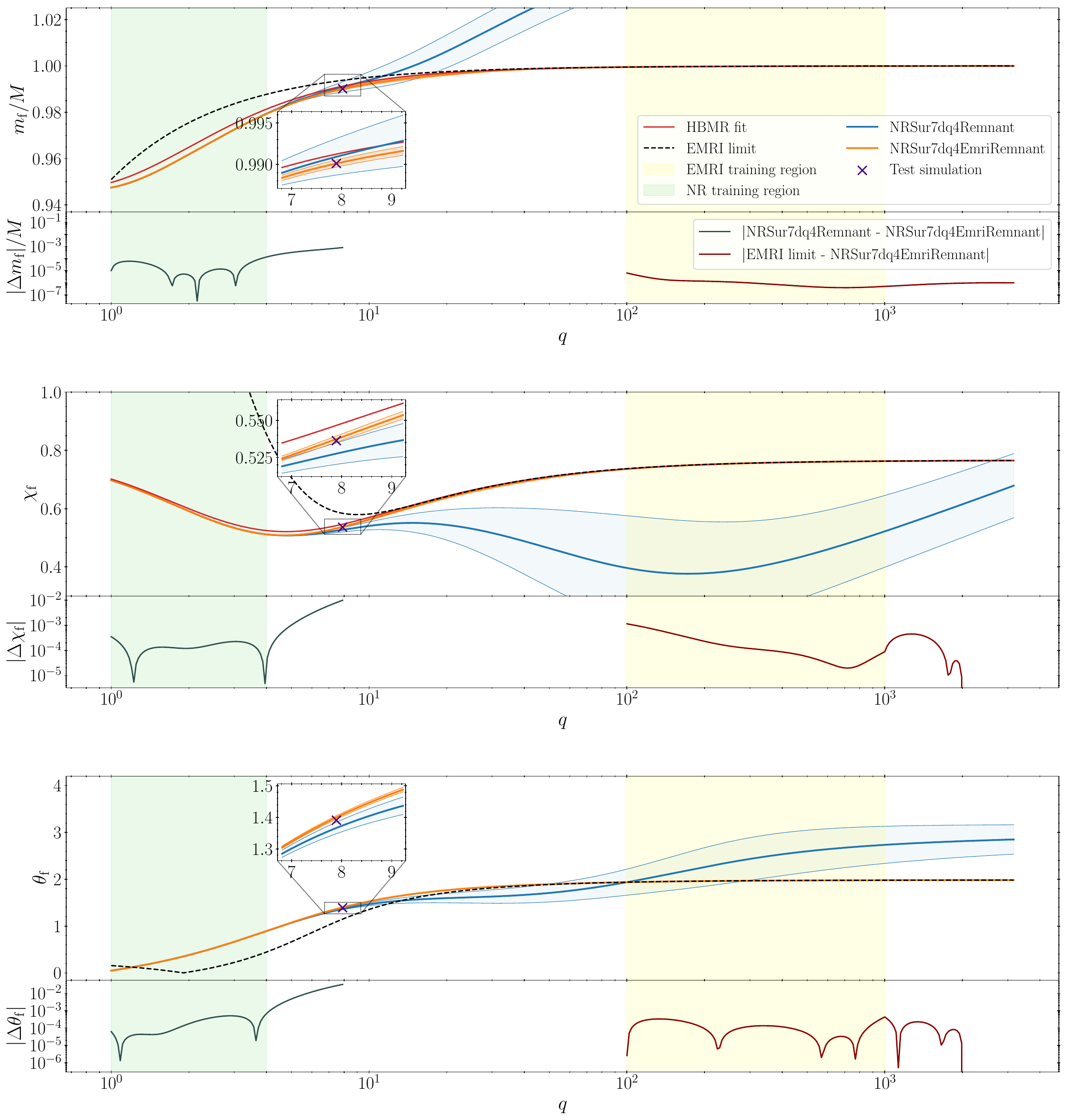}
\caption{
Predictions for the remnant mass $m_{\rm f}/M$ (top panel), spin magnitude
$\chi_{\rm f}$ (middle panel) and spin polar angle  $\theta_{\rm f}$ (bottom
panel). Systems shown in this figure have $\boldsymbol{\chi}_1 = [0.50, -0.49,
-0.31]$, $\boldsymbol{\chi}_2 =[-0.37, 0.37, 0.42]$ and different values of
$q$. Orange, blue, and red curves refer to our extended fits \newname, the previously
published \surname model~\cite{2019PhRvR...1c3015V}, and the HBMR analytic
expressions~\cite{2012ApJ...758...63B,2016ApJ...825L..19H}, respectively. For
the GPR models, solid curves indicate the returned means and shaded areas
indicate 1-$\sigma$ intervals. Black dotted curves show the EMRI limit from
Sec.~\ref{sec:two}. Our model is trained on binaries with mass ratios in the
green (NR) and yellow (EMRI) shaded regions. Purple crosses mark the remnant
properties extracted from a external test simulation that was not used for
either training or internal validation. The bottom subpanels of each panel show
residual  between our model and \surname at low mass ratios and between our
model and the EMRI limit at large mass ratios. %
} 
\label{fig:1dNRTest}
\end{figure*}

An example is shown in Fig.~\ref{fig:1dNRTest}, where we illustrate remnant
predictions for a series of BH binaries with $\boldsymbol{\chi}_1 = [0.50,
-0.49, -0.31]$ and $\boldsymbol{\chi}_2 =[-0.37, 0.37, 0.42]$ %
as a function of
the mass ratio. For these spin values, the test set contains a simulation with
$q=7.90$.

In the NR training region where $q\leq4$,  our new model \newname is essentially
equivalent to the previous version of \surname. The differences between the two
are of are comparable to or smaller than the NR residuals (cf. the
values reported in Fig.~\ref{fig:1dNRTest} against those in Fig. 7 from
Ref.~\cite{2019PhRvR...1c3015V}). At higher mass ratios, the new fit converges
to the imposed EMRI limit with similar residuals. In contrast, the previous
surrogate model departs significantly from the expected limit, in some cases
even returning widely unphysical predictions $m_{\rm f} > M$. %
The new fit
\newname
provides accurate predictions in the intermediate mass ratio region $4<q<100$ even if
we do not have training datapoints. The remnant properties extracted from the
$q=7.90$ test case are within the uncertainties returned by the GPR algorithm;
we report residuals $|\Delta{m}_{\rm f}|\simeq 6.9\times10^{−6} M$,
$|\Delta{\chi}_{\rm f}| \simeq 1.1\times10^{−3}$, and $|\Delta{\theta}_{\rm f}|
\simeq 8.9\times10^{−3}$. The extrapolation at $q > 1000$ is also well behaved;
for the case of the %
final spin, the %
truncation of the residual curve at  $q>
2000$ is a consequence of the regularization procedure described in
Sec.~\ref{sec:three}.

\begin{figure*}[]
\centering
\includegraphics[width=\textwidth]{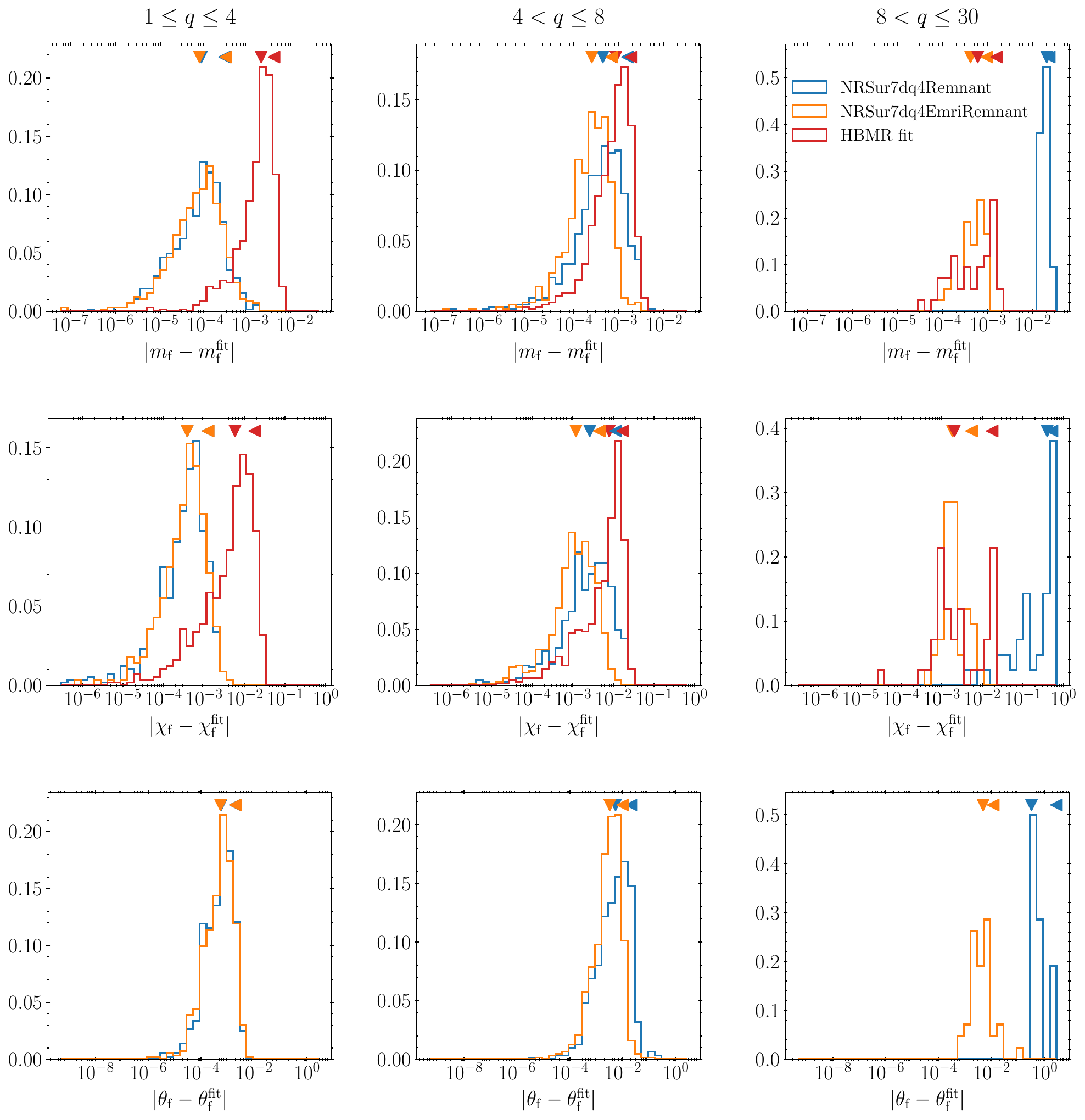}
\caption{
Test residuals for the extended surrogate (orange) against the previous version
of \surname \cite{2019PhRvR...1c3015V} (blue), and the HBMR analytic
fits~\cite{2012ApJ...758...63B,2016ApJ...825L..19H}. Top, middle, and bottom
panels show results for remnant mass $m_{\rm f}/M$, spin magnitude $\chi_{\rm
f}$, and polar angle $\theta_{\rm f}$, respectively. Test simulations are
divided into three mass ratio bins: NR ($q<4$, left), near
intermediate ($4\leq q <8$, middle), and far intermediate ($8\leq q <30$,
right). %
The left- and down-pointing triangles on top
show the $90^{\rm th}$ percentiles and the median for each of the histograms;
the former are also reported in Table~\ref{tableP90}. 
}
\label{fig:NRtest}
\end{figure*}

\begin{table*}
\setlength{\tabcolsep}{5.5pt}
\renewcommand{\arraystretch}{1.5}
\begin{tabular}{
l|l||c|ccc
}

\multicolumn{2}{c||}{} & All  & $1\leq q\leq4$ & $4 < q \leq 8$ & $8 < q \leq 30$ \\\cline{3-6}
\multicolumn{2}{c||}{}   & 100\%  & 45.8\% & 50.7\% & 3.4\% \\
\hline \hline
$m_{\rm f}/M$ & \newname& $6.0 \times 10^{-4}$ & $2.9 \times 10^{-4}$ & $6.9 \times 10^{-4}$ & $9.5 \times 10^{-4}$\\
& \surname &$1.4 \times 10^{-3}$ & $2.7 \times 10^{-4}$ & $1.5 \times 10^{-3}$ & $2.3 \times 10^{-2}$\\
& HBMR fit &$2.8 \times 10^{-3}$ & $3.4 \times 10^{-3}$ & $1.9 \times 10^{-3}$ & $1.6 \times 10^{-3}$\\
\hline
$\chi_{\rm f}$ & \newname& $3.4 \times 10^{-3}$ & $1.3 \times 10^{-3}$ & $4.5 \times 10^{-3}$ & $5.4 \times 10^{-3}$\\
& \surname &$9.3 \times 10^{-3}$ & $1.3 \times 10^{-3}$ & $1.1 \times 10^{-2}$ & $5.3 \times 10^{-1}$\\
& HBMR fit &$1.7 \times 10^{-2}$ & $1.8 \times 10^{-2}$ & $1.7 \times 10^{-2}$ & $1.8 \times 10^{-2}$\\
\hline
$\theta_{\rm f}$ & \newname& $7.3 \times 10^{-3}$ & $2.0 \times 10^{-3}$ & $9.8 \times 10^{-3}$ & $1.2 \times 10^{-2}$\\
& \surname &$1.9 \times 10^{-2}$ & $2.0 \times 10^{-3}$ & $2.1 \times 10^{-2}$ & $2.8 \times 10^0\phantom{-}$\\
\end{tabular}
\caption{
90$^{\rm th}$ percentiles on the test residuals for remnant mass $m_{\rm f}/M$,
spin magnitude $\chi_{\rm f}$, and spin polar angle $\theta_{\rm f}$ computed
for the \newname model presented here, the \surname model from
Ref.~\cite{2019PhRvR...1c3015V}, and the HBMR
fits~\cite{2012ApJ...758...63B,2016ApJ...825L..19H}. The column labeled ``All''
indicates percentiles computed over the entire test set while for the last
three columns we consider subsets of the test simulations in three mass-ratio
bins. The second row indicates the fraction of test simulations in each of
these bins.
}
\label{tableP90}
\end{table*}

\subsection{Test summary}

We now present some summary results using all the NR simulations from the test set.
For convenience, these are separated into three disjoint sets:  %
\begin{itemize}
    \item[(i)]
        ``NR regime'' $1 \leq q\leq4$ (563 simulations). 
        These are simulations covering the same parameter space of \surname
        which were not used to train either that model or ours.
\item[(ii)] ``Near intermediate regime'' $4 < q \leq 8$ (623 simulations). At these moderate mass ratio, the parameter space is reasonably well covered by simulations in the test set~\cite{inprep}.
\item[(iii)] ``Far intermediate regime'' $8 < q \leq 30$ (42 simulations). These NR runs
    are computationally very expensive, resulting in limited number of
    simulations and a sparse coverage of the parameter space. In particular,
    many test simulations in this regime have aligned, non-precessing spins,
    preventing us from fully testing the precession sector. 
\end{itemize}
Note that the largest mass ratio in the test set $q=30$ is still relatively far from the lowest mass ratio $q_{\rm min}=100$ covered by our EMRI training data.

Figure~\ref{fig:NRtest} shows test residuals for $m_{\rm f}/M$, $\chi_{\rm f}$,
and $\theta_{\rm f}$ in each of the tree subsets. The test simulations are
compared against the new model \newname, the previous \surname model from
Ref.~\cite{2019PhRvR...1c3015V}, and the HBMR analytic
fits~\cite{2012ApJ...758...63B,2016ApJ...825L..19H}. The qualitative conclusion
is that our fit has the best performances in all three mass ratio regimes. %
In
particular, it is comparable to the previous \surname model and superior to HBMR when in the NR regime ($1 \leq q \leq 4$). In the near intermediate regime ($4 < q \leq 8$), our results is mildly more accurate than
both  previous fits.  As expected, the previous \surname model behaves poorly
in the far intermediate regime ($8 < q \leq 30$) while including EMRI
information as in this paper returns performance that are similar to, if not
better than, those of the analytical HBMR expressions.
Quantitative results are presented in Table~\ref{tableP90}, where we report the
90$^{\rm th}$ percentile of our test residuals for each of the three mass ratio bins and
well considering the entire test set.

\section{Computational cost}
\label{sec:five}

The computational cost of fitting a GPR model scales as $\mathcal{O}(n^3)$,
where $n$ is the size of the training dataset. On the other hand, evaluating
the fit is an $\mathcal{O}(n^2)$ problem~\cite{RasmussenW06}. Compared to
\surname from Ref.~\cite{2019PhRvR...1c3015V}, the new model \newname requires %
 $n_{\rm EMRI}$
additional binaries and thus take longer to evaluate. We test the performance
of the new model by generating $10^4$ binaries from a broad distribution and
evaluating the time needed to compute $m_{\rm f}$ and $\boldsymbol\chi_{\rm
f}$. The execution times reported below refer to a single thread on an Intel
Xeon Gold 5220R processor.

The previous \surname model requires  $\ssim 2.5 \times10^{-3}$~s to evaluate
$m_{\rm f}$ and $\ssim 7.4\times10^{-3}$~s to evaluate $\boldsymbol{\chi}_{\rm
f}$. The latter is about three times more expensive than the former because the
spin is a vector quantity with three cartesian components while the mass is a
single scalar.
Compared to this baseline, our new  model \newname increases
the computational time by $\ssim 1.2$ times for $m_{\rm f}$ and $\ssim 2.6$ for $\boldsymbol{\chi}_{\rm
f}$. As expected,
this corresponds to a complexity that is roughly quadratic in the size of the
training set $n=n_{\rm NR} +n_{\rm EMRI}$ when considering that $n_{\rm
NR}=1528$ and $n_{\rm EMRI} =250$ (1250). The evaluation time is independent of
the mass ratio and spins, %
with the exception of  the spin fit at $q> 2 q_{\rm max} = 2000$ where we
simply return the EMRI analytical expression.

Overall, %
this additional computational cost is an
acceptable price to pay given the extended parameter space covered by the
augmented model presented in this paper. As the number of available NR
simulations increases and better surrogate models are built, the
$\mathcal{O}(n^2)$ complexity of GPR evaluations will become critical and alternative
regression algorithm will need to be explored.

\section{Conclusions}
\label{sec:six}

We have presented a strategy to augment existing and future NR surrogate models
for the mass and spin of the post-merger BH remnant, extending their regime of
validity to the test-particle limit.  Our approach consists of adding training
data points for binaries with extreme-mass ratios using  analytic  expressions
valid at $\mathcal{O}(1/q)$. 

We applied this procedure to the GPR fit  \surname~\cite{2019PhRvR...1c3015V}
which models precessing binary BHs. We tested our augmentation both internally
via a cross-validation approach (which was also used to select some model
parameters) and externally against a new set of NR simulations. We report
excellent performances: 
\begin{itemize}
\item[(i)] at comparable masses, our new model behaves like the previous \surname from Ref.~\cite{2019PhRvR...1c3015V}  which, in turn,  was shown to be as accurate as the NR simulations used to train it;
\item[(ii)] at extreme mass ratios, our new model reproduces the test-particle analytic limit with similar residuals;
\item[(iii)] in between the two regimes, our new model returns regular and accurate values when compared against test NR runs, outperforming both \surname and the HBMR fits.
\end{itemize}
In summary, we are releasing a single data-driven
model able to  predict post-merger masses and spins across the entire
mass-ratio range, from equal mass binaries to EMRIs. The one drawback is a
moderately higher computational cost; which we quantified and found to be
acceptable given the extended reliability of the model. 

Another limitation of our NR+EMRI approach compared to NR-only surrogates is
the time dependence of the spin evolution. \surname allows users to specify the
time to merger (or orbital frequency) where spins are specified; it then
evolves the spins using the surrogate dynamics to $t=-100M$ where the GPR fits
can be consistently evaluated~\cite{2019PhRvL.122a1101V,2019PhRvR...1c3015V}.
In the absence of a spin-evolution interpolant that captures both comparable
masses and extreme mass ratios, the model presented here can only provide
remnant predictions given pre-merger quantities specified at the reference time
$t=-100M$ before merger (which we approximate with the ISCO in the EMRI case,
cf.  Sec.~\ref{sec:three}). Inputting values (from e.g. GW posterior
distributions) that respect this assumption is left to the user.    

Future work will tackle the modeling of the remnant kick velocity, which was
captured in previous NR-only
surrogates~\cite{2018PhRvD..97j4049G,2019PhRvL.122a1101V,2019PhRvR...1c3015V}.
This is  a more challenging task because BH kicks depend on the orbital phase
at plunge. Building waveform surrogate models spanning the entire mass ratio
range is a considerably more complicated problem and current attempts are
restricted to non-spinning
sources~\cite{2022PhRvD.106j4025I,2020PhRvD.101h1502R}. 

The model presented in this paper, which we dub \newname, is made publicly available through the Python module
\textsc{surfinBH}~\cite{surfin}. More broadly, the augmentation procedure is being integrated in the SXS surrogate codebase and we expect  it
 be valuable for
building future %
BH-remnant models~\cite{inprep}.

\acknowledgements

We thank Costantino Pacilio, Nathan Johnson-McDaniel, and Geraint Pratten for discussions.
M.B. and D.G. are supported by ERC Starting Grant No.~945155--GWmining, Cariplo Foundation Grant No.~2021-0555, MUR PRIN Grant No.~2022-Z9X4XS, and the ICSC National Research Centre funded by NextGenerationEU. 
M.B. is supported by an Erasmus+ scholarship. 
D.G. is supported by Leverhulme Trust Grant No.~RPG-2019-350.
V.V. is supported by European Union Marie Skłodowska-Curie Grant No.~896869.
Computational work was performed at CINECA with allocations through INFN, Bicocca, and ISCRA project HP10BEQ9JB.
This work was supported in part by the Sherman Fairchild Foundation, NSF Grants PHY-2207342, PHY-2011961, PHY-2011968, PHY-2208014, OAC-2209655, AST-2219109, the Dan Black Family Trust, Nicholas and Lee Begovich, the Indian Department of Atomic Energy under project no. RTI4001, and by the Ashok and Gita Vaish Early Career Faculty Fellowship at the International Centre for Theoretical Sciences.

\bibliography{singlemasslimit}

\end{document}